\documentclass[%
 reprint,
 amsmath,amssymb,
 aps,
prb,
]{revtex4-2}

\usepackage{graphicx}
\usepackage{dcolumn}
\usepackage{bm}

\begin{document}

\preprint{APS/123-QED}

\title{Conductivity of high-mobility epitaxial GdN}

\author{Edward X. M. Trewick}
    \email{ted.trewick@vuw.ac.nz}
    \affiliation{School of Chemical and Physical Sciences, Victoria University of Wellington, P.O. Box 600, Wellington 6140, New Zealand.}
    \affiliation{MacDiarmid Institute for Advanced Materials and Nanotechnology, P.O. Box 600, Wellington 6140, New Zealand.}
\author{B. J. Ruck}
    \email{ben.ruck@vuw.ac.nz}
    \affiliation{School of Chemical and Physical Sciences, Victoria University of Wellington, P.O.~Box 600, Wellington 6140, New Zealand.}
    \affiliation{MacDiarmid Institute for Advanced Materials and Nanotechnology, P.O.~Box 600, Wellington 6140, New Zealand.}
\author{W. F. Holmes-Hewett}
    \affiliation{Robinson Research Institute, Victoria University of Wellington, P.O.~Box 33436, Petone 5046, New Zealand.}
    \affiliation{MacDiarmid Institute for Advanced Materials and Nanotechnology, P.O.~Box 600, Wellington 6140, New Zealand.}
\author{H. J. Trodahl}
\affiliation{School of Chemical and Physical Sciences, Victoria University of Wellington, P.O.~Box 600, Wellington 6140, New Zealand.}

\date{\today}

\begin{abstract}
We report electron transport studies of a (001) GdN film grown on the square net presented by the (001) surface of LaAlO$_3$, motivated by recent advances in epitaxial thin-film growth of several lanthanide nitrides. The film we have grown for the purpose is characterised by \textit{in-situ} RHEED and \textit{ex-situ} XRD and XRR to show the best crystallinity and smoothest surfaces we have accomplished to date. It shows a clear ferromagnetic transition at $\sim$~70~K with a saturation magnetisation within uncertainly of 7~$\mu_B$/Gd$^{3+}$~ion, a remanence of 5~$\mu_B$/Gd$^{3+}$~ion and a coercive field of $\sim$~5~mT. It is doped by $\sim$~1\% nitrogen vacancies that introduce $\sim 3 \times 10^{20}$~cm$^{-3}$ electrons into the conduction band. The resistivity shows transport in a conduction band doped to degeneracy by $\sim$0.01 electrons/formula unit with a residual resistance ratio of 2 and a Hall resistivity permitting easily-separated ordinary and anomalous Hall components. The mobility is an order of magnitude larger than we have found in earlier films. 

\end{abstract}

\maketitle

\section{Introduction}

The lanthanide nitrides (\textit{Ln}N) are unique among the lanthanide pnictides for both their ferromagnetic ground state and their controllable semiconductor electron transport~\cite{Natali2013,Holmes-Hewett2020}. Their magnetic moments reside in the \textit{Ln} 4\textit{f} shell, determined by coupled intra-ion spin and orbital moments that precipitate a mixed spin/orbit magnetism. Their exchange interaction within the 5$d$ conduction band ensures that it also shows strong spin splitting. They promise application for cryogenic magneto-electronics, and have already been reported in magnetic Josephson junctions (MJJ)~\cite{Birge2024, Senapati2011, Massarotti2015, Caruso2019, Ahmed2020} and in magnetic cryogenic memory~\cite{Devese_2022,Pot2023,Sharma2024,Cascales2019}. The memory application exploits well-studied magnetism control within the series, but for MJJ it is essential to control also their as yet considerably less well-characterised conductivity. Here we build on recent advances in epitaxial-growth of the $Ln$N~\cite{McNulty2021, Pereira2023,Anton2023, Melendes2024,Tanaka2024, Su2024,Vallejo2024} to advance understanding of their electron-transport in the excellent $Ln$N films that can now be grown.

We select for this initial study the simplest of the series, GdN, with a half-filled 4\textit{f} shell that harbours a purely spin magnetic moment of 7~$\mu_B$ on each Gd$^{3+}$ ion. An early single crystal was described as semi-metallic with electron concentration of $1.9\times10^{21}$~cm$^{-3}$~\cite{Wachter1980,Wachter2016}, while subsequent thin films show an electron concentration that is widely varied over 10$^{15}$ to 10$^{22}$ cm$^{-3}$ controlled by nitrogen-vacancy (V$_N$)~\cite{Melendes2024} and Mg doping~\cite{Natali2013,Lee2015}. It is ferromagnetic (FM) below a Curie temperature of $\approx$~70~K. A spherically symmetric 4\textit{f} orbital state ensures a small coercive field of $<~$10~Oe, all but unique within the \textit{Ln}N series. 

Strongly ionic bonding in the $Ln$N ensures that the valence (conduction) band edge has predominantly N~2\textit{p} (\textit{Ln}~5\textit{d}) character. The strong Coulomb interaction among electrons within the 4\textit{f} shell separates the minority from the majority spin bands that in most of the series leaves 4\textit{f} bands threading the valence (for heavy \textit{Ln}) or conduction (for light \textit{Ln}) bands~\cite{Larson2007}. The largest 4$f$ spin splitting is 12~eV in GdN, leaving the majority-spin (minority-spin) band lying 7~eV below (5~eV above) the gap and in turn removing the 4\textit{f} band from the conduction process. The resulting simplicity of conduction in GdN in part motivates this transport study. 

This is by no means the first magneto-transport investigation of thin GdN films. Earlier studies were, with one exception~\cite{Ludbrook2009}, on less well-ordered films, either non-epitaxial or (111) films grown on hexagonal nets (c-plane sapphire, (0001) AlN) that results in grains of two different in-plane orientation. The past year has seen reports of (001) \textit{Ln}N grown on lanthanum  and yttrium aluminate (LaAlO$_3$, YAlO$_3$), with more uniform crystal alignments~\cite{Pereira2023,Anton2023, Melendes2024,Tanaka2024, Su2024}. Here we report a growth and thorough study of the magneto-transport within such a (001) film, showing most impressively a mobility an order of magnitude larger than the best films grown on hexagonal nets.

\begin{figure}[ht]
    \centering
    \includegraphics[width=\linewidth]{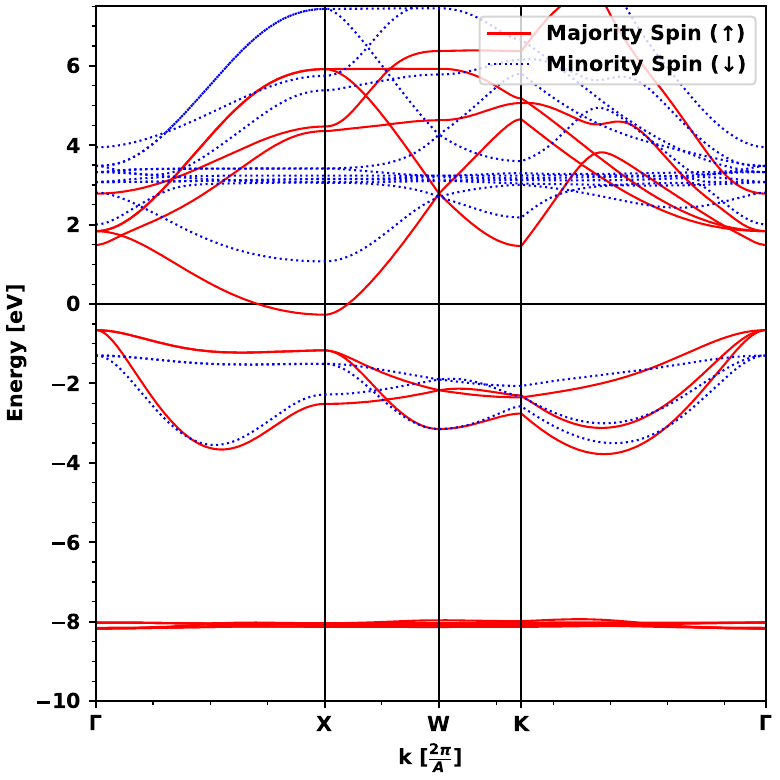}
    \caption{Band structure of GdN, calculated within the generalised gradient approximation to density functional theory with a Hubbard U correction, using a relaxed primitive cell.}
    \label{fig:bandstructure}
\end{figure}

Although stoichiometric GdN is insulating, there is commonly a nonzero n-type conduction resulting from low formation-energy nitrogen vacancy, V$_N$,  defects that are especially prevalent in high-temperature-grown films~\cite{Punya2011,Porat2024}. Each defect releases three electrons, of which computations suggest two are trapped by localised V$_N$ states, leaving one in a conduction band that in GdN is otherwise altered minimally by a low V$_N$ concentration~\cite{Holmes-Hewett2025}. A density functional theory GGA+\textit{U} band structure computed in the ferromagnetic ground state is shown in Figure~\ref{fig:bandstructure}. Fermi level is shifted to be zero at a band filling of 1 electron per 100 Gd$^{3+}$ ions (1\% V$_N$, a carrier concentration of $3.2\times10^{20}\,\text{cm}^{-3}$), and shows that 4\textit{f}/5\textit{d} exchange and thus spin splitting is large enough that only the majority-spin band is occupied at carrier concentrations below $\sim10^{21}\,\text{cm}^{-3}$. Well below this level of doping the conduction band is occupied to degeneracy, resembling a half metal. There are three equivalent prolate-spheroid electron pockets centred at X points with a semi-major axis along $\Gamma$-X that is a factor of $\sim~3$ larger than in the semi-minor transverse (X-W) direction, in turn signalling a factor $\sim$~10~effective mass contrast~\cite{Holmes-Hewett2025}. Below we use this band structure to simulate the magnetoconductivity as a comparison with the measured results.

The next section describes the growth specifics, the structural characterisation and the magnetic/magnetoconductivity measurements. Section III reports structural and magnetic characterisation results, with the transport results and their interpretation in Section IV. Our conclusions are summarised in Section V. 



\section{Experimental details}

The films used in this study were deposited on (001) LAO in the presence of high-purity (99.9995\%) N$_2$ gas, relying on the spontaneous decomposition of N$_2$ on the clean Gd surface to form the nitride~\cite{Ullstad2019}. Films were deposited in a Thermionics MBE chamber with a base pressure of $5\times10^{-9}\,\text{mbar}$. The substrates were outgassed for 2 hours at $\sim500^\circ$C, and then held at $\sim400^\circ$C during growth. Gadolinium metal was evaporated using an electron beam at a nominal rate of $0.02\,$nm/s in the presence of 10$^{-4}$~mbar molecular nitrogen. We routinely expect that GdN films grown under these conditions are doped to V$_N~\sim$~0.01 of the nitrogen sites. 

Reflection High Energy Electron Diffraction (RHEED) was performed \emph{in situ} during deposition to confirm in-plane epitaxial orientation.
The films were then capped with a 60~nm passivation layer of aluminium nitride (AlN) to reduce oxidation. Further structural, magnetic, and electrical characterisation was performed \textit{ex situ}. After removal from the chamber exposure to air was minimised, though several hours of exposure during measurement is inevitable. Care was taken to inspect the films regularly, checking for serious destructive reactions.

Structural information was determined from symmetric $2\theta/\omega$ scans, X-Ray Reflectivity (XRR), and rocking curves using a $\chi-\phi$ goniometer stage on a 2-circle Rigaku SmartLab X-ray diffractometer. Magnetic susceptibility measurements were made from 5-300~K, along with FM hysteresis loops at selected temperatures in fields up to $\pm\,7\,T$. Magnetic measurements were conducted in a Quantum Design MPMS-XL SQuID magnetometer for fields applied in three crystallographic orientations; [100] and [110] in-plane, and [001] out-of-plane. 

For electrical measurements 0.5~nm~Ti/50~nm~Au contacts were deposited onto the AlN capping layer, and two Hall bars with dimensions w/l of 0.4/3.4~mm were subsequently formed lithographically by argon ion milling defined with an AZ1518 photoresist.

The Hall bars were formed with the current directions along  the (100) and (110) in-plane directions to measure the magnetoresistivity in these directions. The contact resistances were $\lesssim$~500~$\Omega$. Resistivity measurements were performed on the two Hall bars in a Quantum Design PPMS DynaCool across the full temperature range of 300-1.8~K, and more detailed longitudinal and transverse magnetoresistance studies at 1.8~K. The results on the two bars were identical within uncertainty.

\section{Structural and magnetic characterisation}

\begin{figure}[ht]
    \centering
    \includegraphics[width=0.9\linewidth]{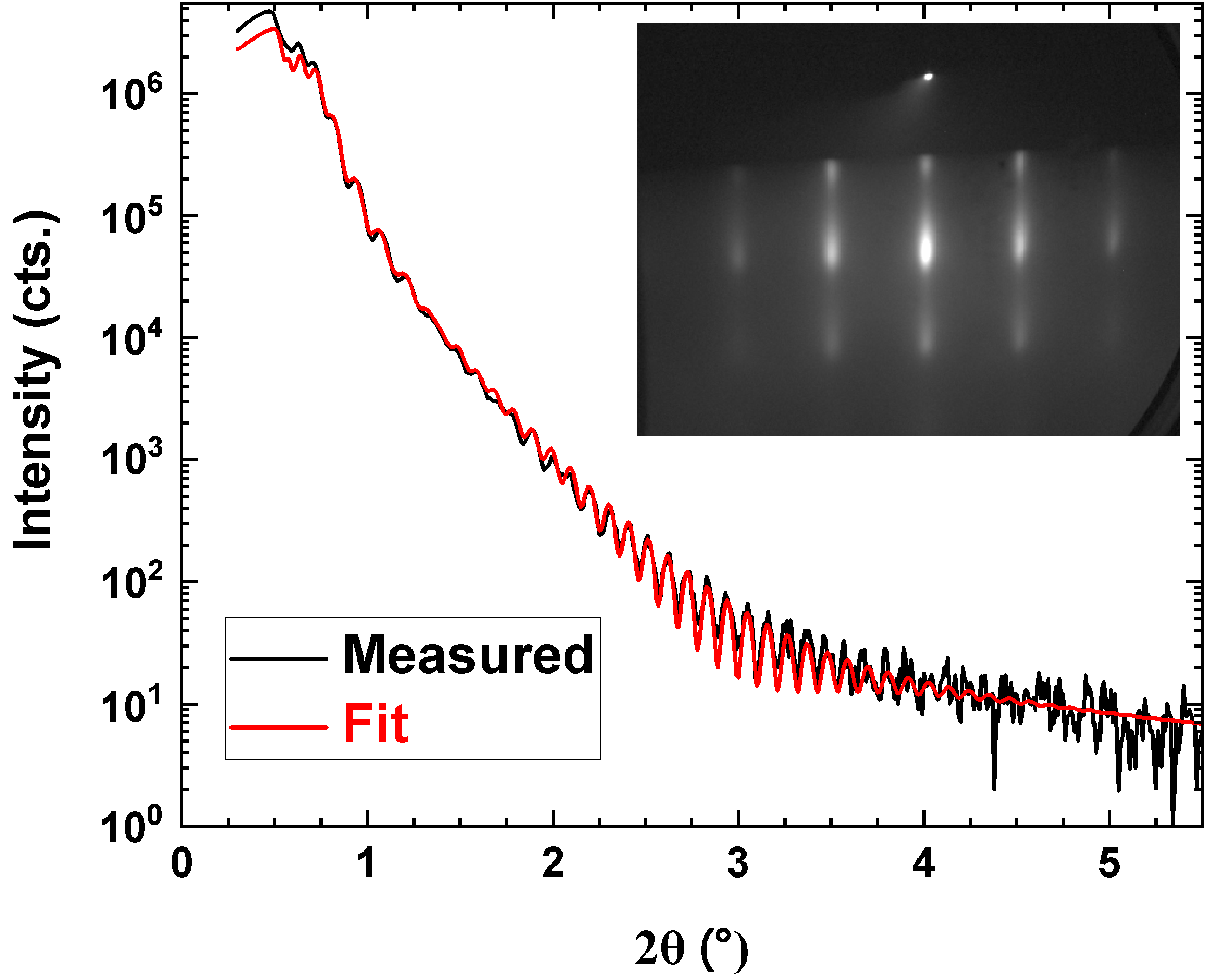}
    \caption{X-Ray Reflectivity (XRR) data (blue) fitted with GenX (black), showing the 56nm, 2.6nm RMS roughness AlN layer and 79.4nm, 0.82nm RMS roughness GdN layer. The inset is a RHEED image from the end of the growth, further supporting the smooth top (001) surface.}
    \label{fig:XRD_xrr}
\end{figure}

The RHEED image of the GdN film before capping is seen in the inset of Figure~\ref{fig:XRD_xrr} and shows a streaky pattern signalling a relatively smooth surface. The pattern is comparable to the reported epitaxial growths on the same substrate of HoN and SmN~\cite{Pereira2023, Melendes2024}. XRD showed clear features corresponding to GdN (002) and (004) reflections signalling a lattice constant of $4.983(1)\,\text{\AA}$, with a (002) rocking-curve width of $1.395(5)^\circ$, and a phi scan confirmed the expected epitaxial alignment; GdN (100) parallel to LAO (110). X-ray reflectivity data (XRR) in Figure \ref{fig:XRD_xrr} yielded the layer thickness of 79.4~nm GdN (roughness 0.8 nm) GdN capped by a 56~nm (roughness 2.6 nm roughness) AlN layer.



\begin{figure}[ht]
    \centering
    \includegraphics[width=0.9\linewidth]{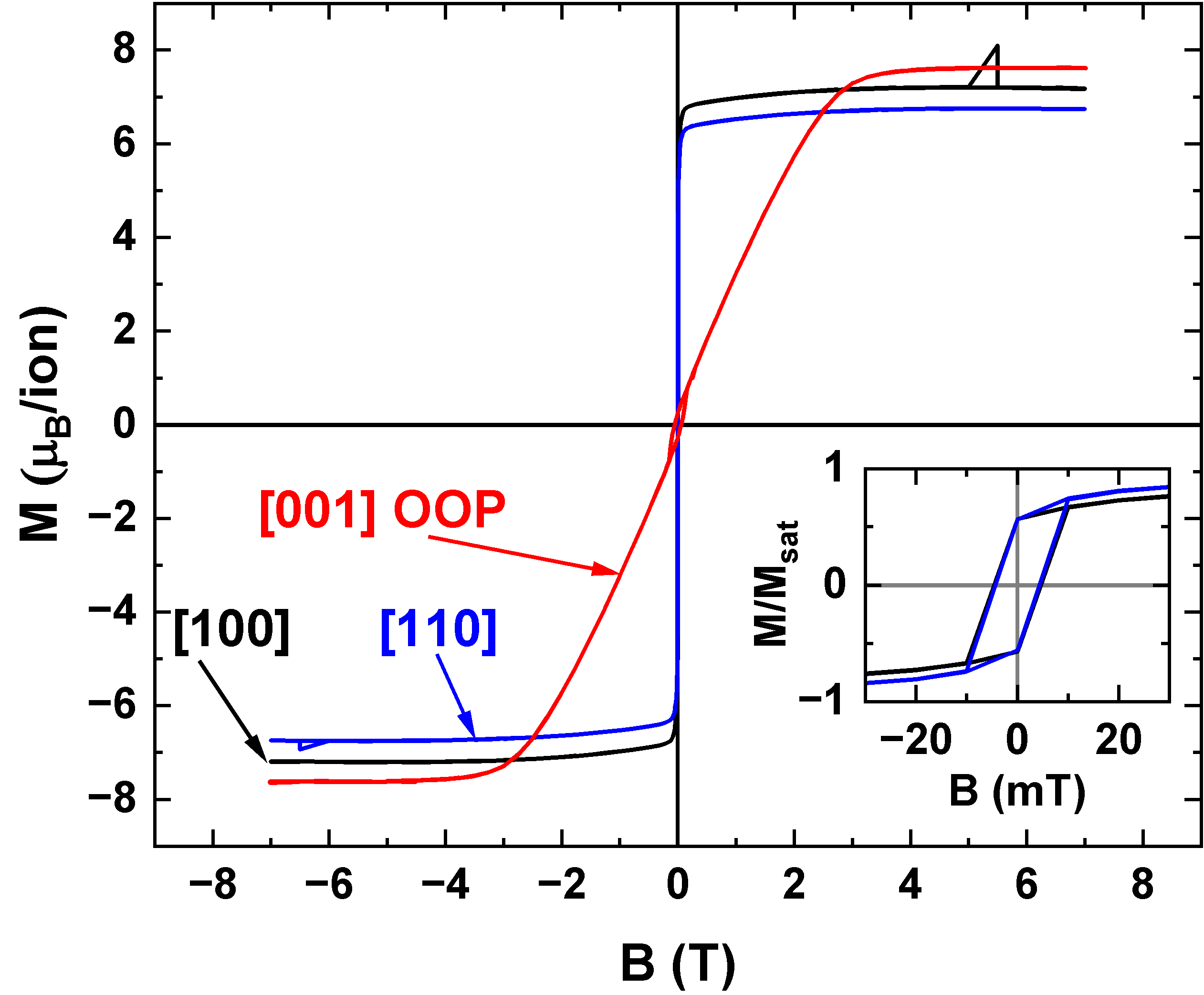}
    \caption{Magnetisation hysteresis loops at 5 K with hi-res inset. The two in-plane traces (black, blue) show the same behaviour, except for a few \% scale factor associated with the sample area uncertainty. The (red) field out-of-plane data show the usual nearly linear rise between the fields of $\pm$4$\pi$M.}
    \label{fig:SQuID_MH_5K}
\end{figure}

Magnetisation was measured on two rectangles of the films cut to be square with each of the $\langle 100 \rangle$ and the $\langle 110 \rangle$ directions.
The temperature dependence showed a ferromagnetic onset at a Curie temperature of $\approx$~70~K. The 1.8~K hysteresis traces in Figure \ref{fig:SQuID_MH_5K} show the same saturation magnetisation 7~$\mu_B$ per Gd$^{3+}$ ion, as is typically found in GdN, within the uncertainty of the sample volume. The field-in-plane magnetisation, shown expanded in the inset, show a coercive field of a few mT and a remanence of $\sim$~5~$\mu_B$ per Gd$^{3+}$ ion, 70\% of the saturation magnetisation. In contrast the field-normal magnetisation displays typical shape anisotropy, rising linearly toward saturation only as the applied field reaches 4$\pi$M~$\approx~2.6$~T. The small open hysteresis suggests that the field was $\sim$0.5$^\circ$ out of exactly normal, which drives an in-plane magnetisation well below the shape anisotropy field.

\section{Magnetoconductivity}

\begin{figure}[ht]
    \centering
    \includegraphics[width=0.9\linewidth]{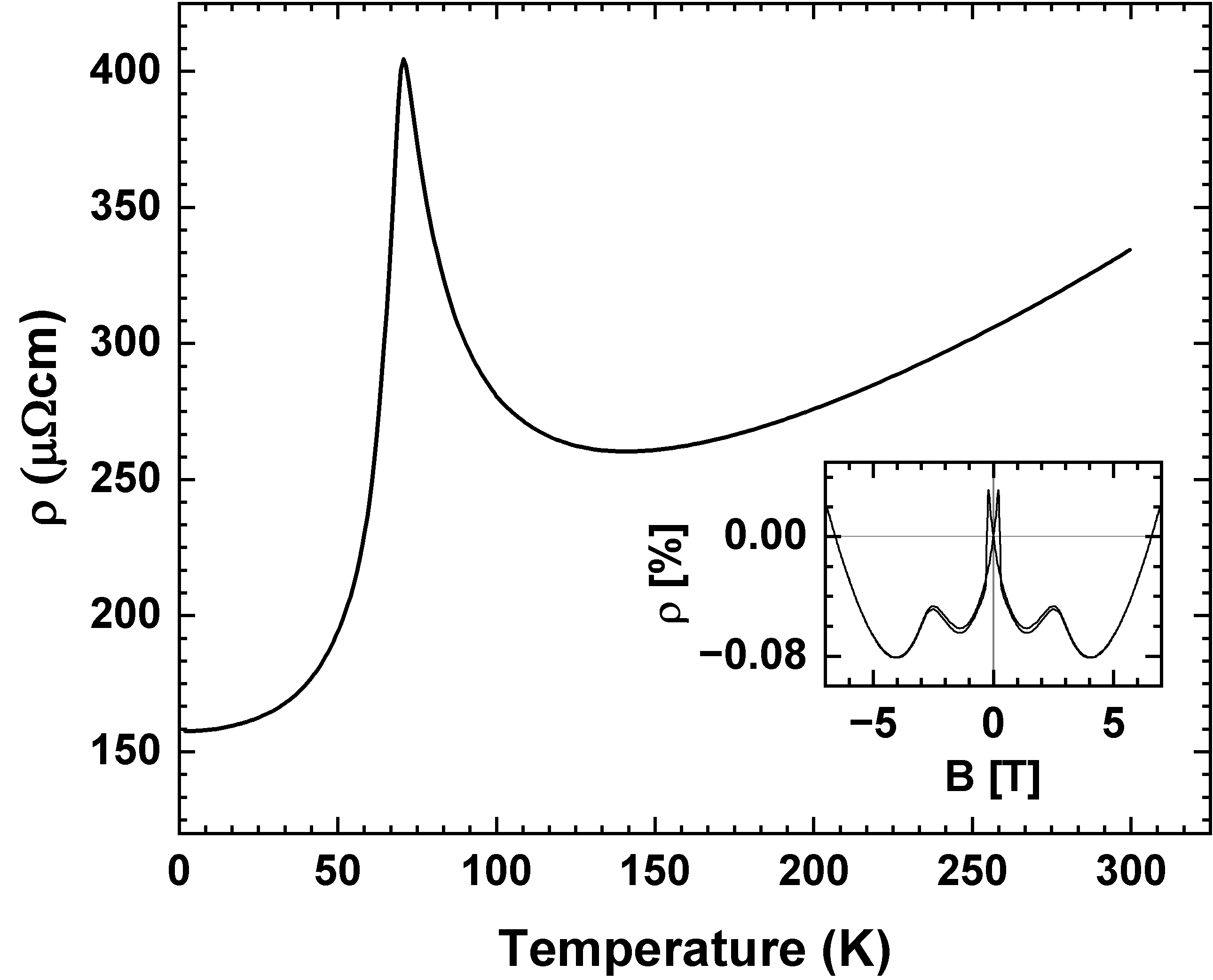}
    \caption{Temperature dependence of the resistivity in zero applied field. The field-out-of-plane magnetoresistance, [$\rho$(B)-$\rho$(0)]/$\rho$(0) is plotted in the inset.}
    \label{fig:PPMS_TDR}
\end{figure}

 Figure~\ref{fig:PPMS_TDR} shows the temperature-dependent resistivity of the [110] aligned Hall bar in zero magnetic field. As is required by symmetry the [100] bar showed the same resistivity. The curve is dominated by magnetic-disorder scattering peaking near $T_C$ and otherwise there is a general positive temperature coefficient of resistivity typical of transport in a band of extended states found in metals and in semiconductors doped to degeneracy. As reported earlier in epitaxial films~\cite{Maity2020}, the magnetic-disorder resistivity is seen to fall to zero deep in the FM phase well below T$_C$ and become weak and temperature independent well above T$_C$, and is strongly diminished by a magnetic field of a few Tesla. Figure \ref{fig:PPMS_TDR} shows a linear temperature dependence above $\sim$200~K that is consistent with phonon scattering, which extrapolates toward the residual zero-temperature resistivity found below 10 K. The residual resistance ratio $\rho$(300~K)/$\rho$(1.8~K)~$\approx$~2.0 signals residual defect scattering equal to phonon scattering at 300~K.
 
Applying an out-of-plane magnetic field deep in the FM phase (5~K) we find a small $\leq 0.1$\% magnetoresistance shown in the inset of Figure \ref{fig:SQuID_MH_5K}. It is positive at fields above 5 T but with weak features at both $\mathrm{B}=\pm 4\pi$M, the saturation field for the geometry, and at the out-of-alignment in-plane fields where hysteresis is seen in Figure \ref{fig:SQuID_MH_5K}. There is no suggestion at 5~K of the negative MR that follows from a field-driven diminishing of magnetic-disorder scattering that is strong close to T$_C$~\cite{Maity2020}. 

\begin{figure}[ht]
    \centering
    \includegraphics[width=\linewidth]{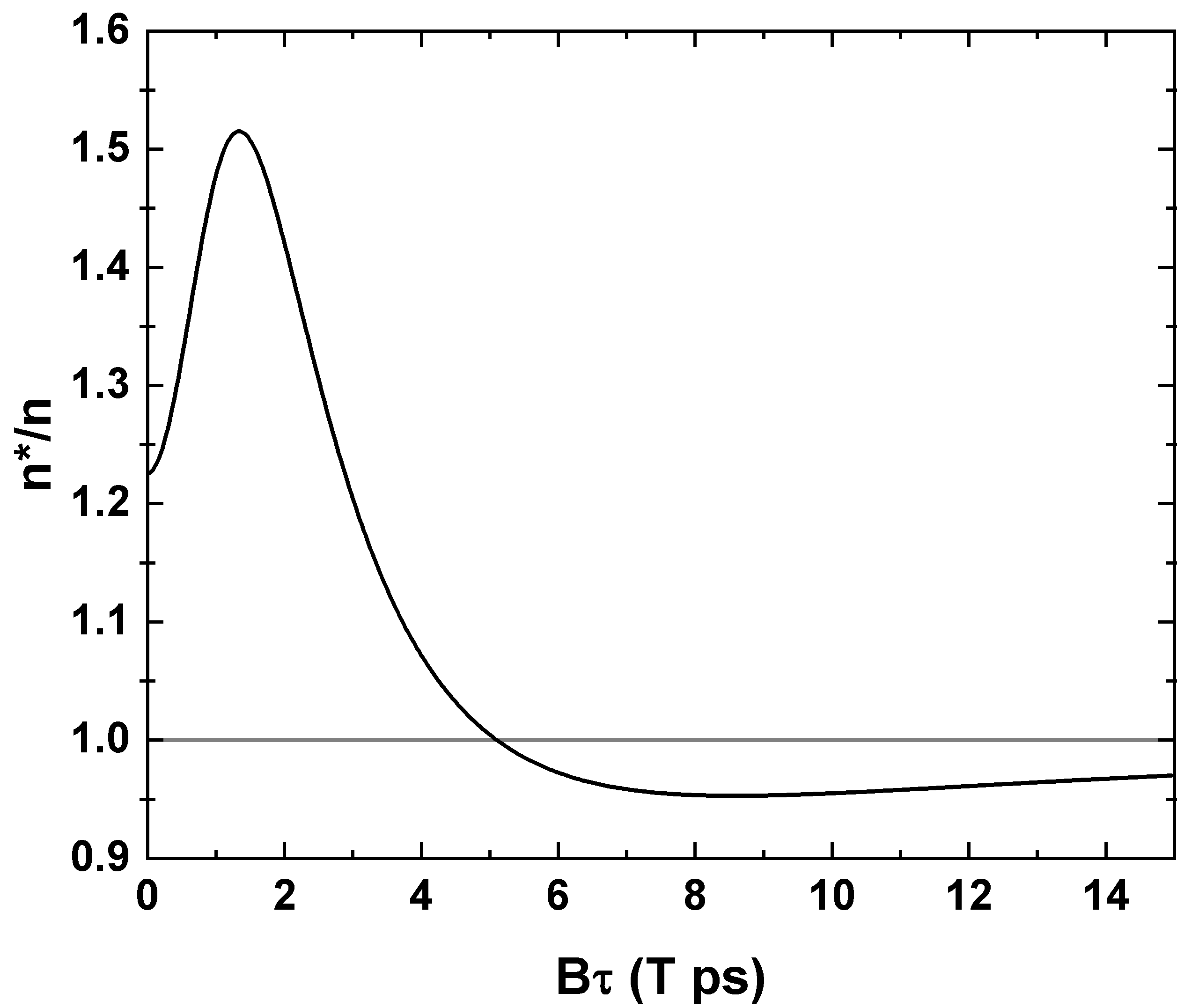}
    \caption{``Carrier concentration'' as determined by the standard formula used for measurements of the ordinary Hall effect (OHE), here using the $\rho_{xy}$ element of the analytically derived resistivity tensor for the Fermi surface of GdN. The dashed line shows the actual actual carrier concentration, to which $\frac{1}{e}\frac{\mathrm{d}B}{\mathrm{d}\rho}$ tends for large $B\tau$}
    \label{fig:Analytical_Btau_n}
\end{figure}

The Hall resistance is linear in field at ambient temperature over the full field range up to 7~T. Within the usual procedure it yields a carrier concentration $3.2\times 10^{20}$~cm$^{-3}$. The Hall slope shows no significant temperature dependence, in combination with the resistivity giving a remarkably high electron mobility of 130~cm$^2$/Vs at 1.8 K. This high mobility implies the film is free of disorder-induced localisation effects, however the conventional model relating the Hall coefficient to the carrier concentration still relies on conduction in a simple quadratic and spherically symmetric band, which must be questioned for conduction in the small ellipsoidal carrier pockets shown in Figure~\ref{fig:bandstructure}. For reassurance in this regard we have exploited a Boltzmann-equation treatment~\cite{mackeyMagnetoconductivityFermiEllipsoid1969} to calculate the field-dependent Hall resistivity from the set of three ellipsoidal pockets in total harbouring $\mathrm{n} = 3.2\times 10^{20}$~cm$^{-3}$ electrons. Using the local slope d$\rho_{Hall}$ to determine a Hall-estimated carrier concentration (n*), shown plotted as normalised to the real concentration in the three spheroids in Figure \ref{fig:Analytical_Btau_n}. On this scale the available fields are very close to zero, suggesting that our data overestimate the carrier concentration by $\sim$~20\%, leading in turn to a 20\% \textit{underestimate} of the electron mobility. 

\begin{figure}[ht]
    \centering

    \includegraphics[width=0.9\linewidth]{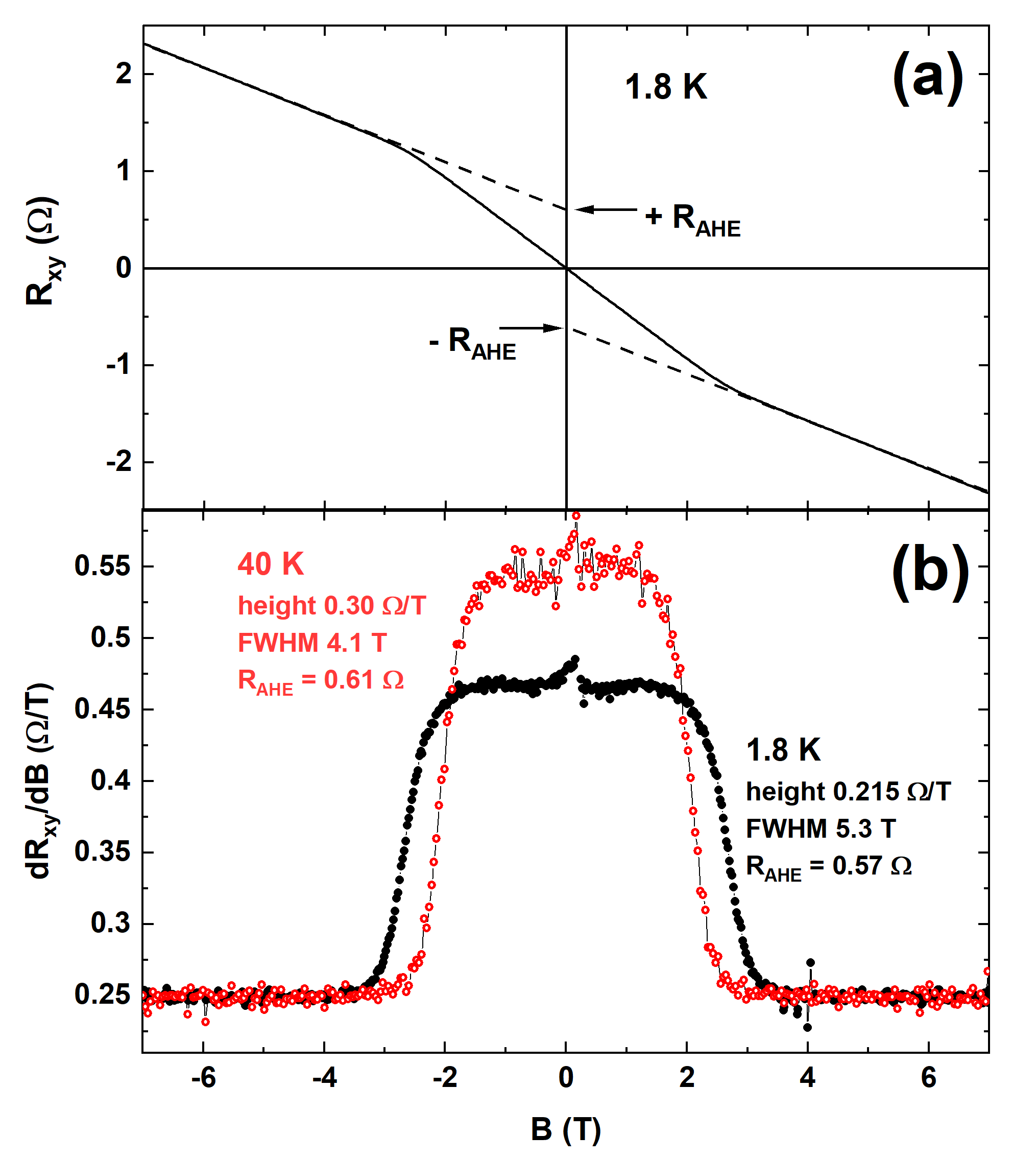}
    \caption{(a) Hall resistance at 1.8~K deep in the FM phase. At fields above 2.6 T it is purely linear, but with an AHE offset characterised as $\pm$R$_{\mathrm{AHE}}$ across $\pm$2.6~T. This low-temperature pattern is found across the entire FM phase, with very similar offsets but quite different field width where the offset develops. (b) The local slope of the Hall resistance from (a) along with similar results at 40~K. The constant high-field slopes are the ordinary Hall effect, yielding the carrier concentration, and the area under the hat-enhancement is the AHE offset, 2R$_{\mathrm{AHE}}$.}
   \label{img:AHEtimes2}
\end{figure}


Below the Curie temperature the Hall resistivity shows a contribution from a transverse spin-orbit force acting on the moving electrons' magnetic moments. The resultant anomalous-Hall-effect resistance R$_{\mathrm{AHE}}$ follows the spin imbalance in the transport channel that in the Hall geometry is proportional to the out-of-plane magnetisation seen in red in Figure \ref{fig:SQuID_MH_5K}~\cite{Trodahl2017}. In the field out-of-plane Hall geometry it is linear in field up to B=4$\pi$M and stably saturated at higher fields. The result is a piecewise linear R$_{AHE}$ with an enhanced slope between applied fields $\pm$ $4\pi$M. Exactly this pattern is evident in the Hall resistance data of Figure \ref{img:AHEtimes2} (a), where the ordinary Hall slope appears beyond $\pm$ 4$\pi$M. Between those limits the slope is enhanced by R$_{AHE}$/4$\pi$M.



A clearer interpretation of these results is offered by plotting the derivative of the data in Figure \ref{img:AHEtimes2}~(a) as the field is varied across the available $\pm$~7~T range, as seen in Figure \ref{img:AHEtimes2}~(b). In the high-field region the slope is constant, but shows a hat-like enhancement in the range $\pm$~2.6~T. The width of the hat is gives the saturation magnetisation directly, here $4\pi\text{M}\approx 2.65\;\text{T}$ at 1.8~K, reduced to 2.05~T at 40~K. This implies a magnetisation of $\sim7.1\;\mu_B/\text{ion}$ at $1.8\;\text{K}$, reducing to $\sim5.5\;\mu_B/\text{ion}$ at $40\;\text{K}$, consistent with the magnetometry data.  On the other hand, the the slope enhancement is R$_{AHE}$/4$\pi$M, so that the area under the hat is 2R$_{AHE}$, a direct measure of the AHE offset, which is in turn proportional to the spin imbalance in the transport band. We thus use the full width at half maximum (FWHM) as an estimate of 
8$\pi$M and the product of the FWHM and the slope difference between low and high fields as R$_{AHE}$, with the results seen in Figure~\ref{img:AHEtimes2}. The reduced M$_{sat}$ with rising temperature is common among FM materials, though the $\sim$37\%  between 1.8 and 40 K is relatively strong. The lack of such a fall in the conduction-band spin imbalance signals that the minority-spin electron pockets remain vacant even at the reduced magnetisation. Note that the apparent $\sim$7\% spin-imbalance rise at 40 K is only marginally within our estimated uncertainty. 
In the context of itinerant ferromagnetism of such familiarity in metallic ferromagnets one might wonder that the carrier spin imbalance remains fixed, perhaps even rising as the saturation magnetisation falls from 1.8 to 40~K. Clearly the spin splitting at the CB minimum remains sufficient to leave the minority-spin CB empty even at 40 K. Note that here a fully spin-aligned pocket with $\sim$~1\% occupation corresponds to a magnetisation enhanced by only 0.01~$\mu_B$ per Gd$^{3+}$ion, almost three orders of magnitude smaller than the 7$\mu_B$ 4$f$ magnetisation and well below the sensitivity of the magnetisation measurement. 

\section{Summary}

This paper describes an electron-transport study of an epitaxial GdN film with uniform (001) alignment in a strong contrast with structural domains common in epitaxial (111) films grown on hexagonal nets. It was characterised by RHEED, XRD and XRR to show excellent crystal quality, smooth surfaces and uniform thickness. Its saturation magnetism is 7 $\mu_B$/Gd$^{3+}$, with a coercive field of a few mT and a 70\% remanence, excellent for its use as a soft magnetic component in tri-layer memory devices. The film has a conduction band doped to degeneracy by $\sim$ 1\% nitrogen vacancies and an electron mobility of $\sim$130~cm$^2$/Vs, a factor of ten improvement in comparison with (111) films. The residual resistance ratio, $\rho$(300K)/$\rho$(5K) $\sim$~2 is the largest we have measured in electron-doped GdN. The Hall effect in the ferromagnetic phase shows a clear pattern allowing separation of the ordinary and anomalous Hall contributions permitting its use to determine both the saturation magnetisation and the spin imbalance in the conduction-band channel. 

\section{Acknowledgements}

This research was supported by Quantum Technologies Aotearoa (contract UOO2347), a research programme
of Te Whai Ao — the Dodd Walls Centre, and by the New Zealand Endeavour fund (contract RSCHTRUSTVIC2447). Travel support was received from 
a QuantEmX grant GBMF9616 from ICAM and the Gordon and Betty Moore Foundation. The MacDiarmid Institute is supported under the New Zealand Centres of Research Excellence programme. The authors have benefited enormously from detailed discussions with Jackson Miller, Simon Granville, and Bob Buckley.

\bibliography{GdNtransport}

\end{document}